\documentclass[twocolumn,showpacs,preprintnumbers,amsmath,amssymb]{revtex4}

\usepackage{graphicx}
\usepackage{dcolumn}
\usepackage{bm}

\begin{document}

\title{Global Optimization of Minority Game by Smart Agents}
\author{Yan-Bo Xie$^{1}$}
\author{Bing-Hong Wang$^{1}$}
\email{bhwang@ustc.edu.cn, Fax:+86-551-3603574.}
\author{Chin-Kun Hu$^{2}$}
\author{Tao Zhou$^{1}$}
\affiliation{%
$^{1}$Nonlinear Science Center and Department of Modern Physics,
University of Science and Technology of China, Heifei, 230026, PR
China \\
$^{2}$Institute of Physics, Academia Sinica, Nankang, Taipei
11529, Taiwan
}%

\date{\today}

\begin{abstract}
We propose a new model of minority game with so-called smart
agents such that the standard deviation $\sigma^2$ and the total
loss in this model reach the theoretical minimum values in the
limit of long time. The smart agents use trail and error method to
make a choice but bring global optimization to the system, which
suggests that the economic systems may have the ability to
self-organize into a highly optimized state by agents who are
forced to make decisions based on inductive thinking for their
limited knowledge and capabilities. When other kinds of agents are
also present, the experimental results and analyses show that the
smart agent can gain profits from producers and are much more
competent than the noise traders and conventional agents in
original minority game.
\end{abstract}


\maketitle

\section{Introduction}

The minority game (MG) models was introduced by {\it Challet} and
{\it Zhang} in 1997 as a model for the competition for limited
resources\cite{Zhang 97}, which have attracted much attention in
recent years. The basic scenario is easy to explain: there is a
population of $N$ players who, at each time step, have to choose
either 0 or 1. Those who are in the minority win, the other lose
(to avoid ambiguities, $N$ is chosen to be odd). The agents make
their decisions based on the most recent $m$ outcomes, thus there
are $2^m$ different histories. A strategy is defined as a table of
$2^m$ choices (either 0 or 1) for the $2^m$ corresponding
histories, so that there are $2^{2^m}$ different strategies in the
strategy-space. Each agent randomly picks $s>1$ strategies from
the strategy-space in the beginning of the MG. To each strategy is
associated a integral point, which initially takes the value 0 and
will increase by 1 at each time step if it predicts the result
correctly. Each agent uses the one with the highest point among
his $s$ strategies, if there are several strategies with the same
highest point, one of those will be chosen randomly. A very
important quantity in this model is the overall loss defined as
\begin{equation}
L(t)=N_{{\rm loss}}(t)-N_{{\rm win}}(t)\geq 1
\end{equation}
where $N_{{\rm loss}}$ and $N_{{\rm win}}$ are, respectively, the
number of losers and the winners at time $t$. Apparently, the
smaller $L(t)$ is, the better the system performs. Another related
quantity is called the standard deviation and defined as
\begin{equation}
\sigma^2(t)=(n_0(t)-\bar{n})^2
\end{equation}
where $n_0$ is the number of agents who choose 0 and
$\bar{n}=N/2$. It is easy to see that $\sigma^2(t)=L^2(t)/4$ and
theoretically, the minimum value of $\sigma^2(t)$ is 0.25.

One of the focuses of scientists' attention is the problem how to
improve the performance of system, i.e. to reduce $\sigma^2$.
Recently, some new kinds of agent are introduced\cite{Savit
00,Wang 02}, by whom the overall performance of system is
improved. A farther question is whether it is possible to achieve
the global optimization in the framework of the MG model assuming
that agents try to outsmart each other for their selfish gain and
act based on inductive thinking\cite{Inductive}.

Recently, a significant work is achieved by {\it Reents}, et al,
who propose a stochastic minority game model in which $\sigma^2$
is minimized\cite{Reents 01}. In their model, an agent will not
change his choice in the next time step if he wins in the present
turn, by contraries, he will change his choice at probability $p$.
The value of $p$ is the same for all the agents. When $p\ll 1/N$,
{\it Reents} et al, found that $\sigma^2\sim 0.25$. However, the
agents in real-life systems are not as clever as {\it Reents},
they do not know how to select a value of $p$, and even do not
know the total number of agents $N$. Thus {\it Reents}'s model may
be not proper for the systems consisting of agent with inductive
thinking.

{\it Metzler} and {\it Horn} have introduced the evolution into
the stochastic minority game model\cite{Metzler 03}.  Similarly to
the evolutionary minority game model\cite{EMG}, for an arbitrary
agent {\bf i}, a probability $p_i(t)$ and a score $s_i$ is
equipped\cite{ex1}. The score $s_i$ increases by 1 if the agent
wins and decreases by 1 if the agent loses. When $s_i\leq d<0$,
the agent is deceased and replaced by a new agent with a reset
score $s_i=0$. If $p_i(t)$ of the new agent is randomly
distributed in $(0,1)$, the average value of $p_i(t)$ in the final
stationary state is found to be at the order of 1 and thus
$\sigma^2\sim O(N^2).$ They also discussed the situation in which
the new agent chooses $p_i(t)$ by copying the value of $p_j(t)$ of
another agent who is randomly selected. Within this scheme, it is
possible to see that $p_i(t)\sim O(1/N)$ and $\sigma^2\sim 1$ in
sufficiently long time. However, it is still unreasonable to
assume that an agent knows the information of all other agents.
Furthermore, $p$ is at the order of $1/N$ and thus $\sigma^2$ is
greater than 0.25 in the final state. The best solution is still
not achieved in their model.

In the present paper, we propose a new model of minority game with
so-called smart agents such that the standard deviation $\sigma^2$
and the total loss in this model reach the theoretical minimum
values in the limit of long time. The smart agents act based on
inductive thinking but bring global optimization to the system.
Experimental results and analyses show that when other kinds of
agents are also present, the smart agent can gain profits from
producers and are much more competent than the noise traders and
conventional agents in original minority game.

\section{Model and Numerical Simulation}

Our model consists of $N$ agents with $N$ an odd integer. Each
agent has only one strategy which evolves with the following rule:
suppose at a given time step $t$, the memory (history) is $\mu$
and the strategy of the $i$-th agent is $s_i(t,\nu)$ for
$\nu=0,...,2^M-1$. Also, each agent has a probability function
$p_i(t,\nu)$ for $i=1,...,N$ and $\nu=0,...,2^M-1$. If the $i$-th
agent wins at $t$, the strategy will not be changed; contrarily,
with probability $1-p_i(t,\mu)$, $s_i(t,\nu)$ is not changed, with
probability $p_i(t,\mu)$, $s_i(t+1,\mu)=1-s_i(t,\mu)$, but
$s_i(t+1,\nu)=s_i(t,\nu)$ for all other $\nu\ne\mu$.

The initial value of $p_i(t,\nu)$ is randomly selected in $(0,1)$
and evolves by self-teaching mechanism, which is the simplest
trail and error method. For a given time step $t$ with history
$\mu$, consider the last time step $t'$ when the memory is also
$\mu$. If the agent {\bf i} won at $t'$ or he loses but does not
change $s_i(t',\mu)$, then no changes will occur. Otherwise,
$p_i(t,\mu)$ will change according to the following
rule\cite{ex2}:
\begin{equation}
p_i(t+1,\mu)=\left\{
    \begin{array}{cc}
        {\rm min}(1,2p_i(t,\mu)) &\mbox{agent {\bf i} wins at time $t$}\\
        p_i(t,\mu)/2 &\mbox{agent {\bf i} loses at time $t$}
    \end{array}
    \right.
\end{equation}
No changes will occur for all $p_i(t,\nu)$ with $\nu \neq \mu$.

Note that the evolution of $p_i(t,\mu)$ for different memories is
essentially decoupled in our model.  Therefore, mathematically
speaking, the $m\ne 0$ case is a trivial generalization of the
$m=0$ case. The reason why we introduce different memories here is
to mimic the behavior of the agents in real-life markets that the
agents study the selection rules for different memories in order
to find some regularities.

\begin{figure}
\scalebox{0.4}[0.4]{\includegraphics{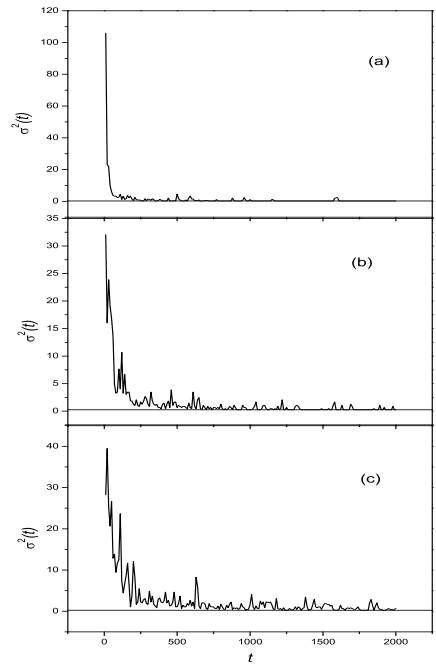}}
\caption{\label{fig:epsart} Time evolution of $\sigma^2 (t)$ for
$N=101$ smart agents with $m=$0 (a), 1 (b), and 2 (c). The value
of $\sigma^2 (t)$ shown in this figure is the average of 10
independent experiments and the horizontal line represents
$\sigma^2=0.25$.}
\end{figure}

Figure 1 shows the simulation results, which indicate that the
system will reach global optimization in sufficiently long time.
We have checked that the property of time evolution of
$\sigma^2(t)$ for the cases with more agents and larger memory is
the same as that of $N=101$ and $m=0,1,2$.

Fig.2 presents the log-log plot for the time dependence of
$G(t)=\sum_{i=1}^N p_i(t)$ and $H(t)=\prod_{i=1}^N p_i(t)$ for
$N=101$ and $m=0$, respectively. The results show that $G(t)$ has
a power law dependence of time with the exponent $\gamma \approx
-1$ when $t$ is large, which suggests that $G(t)\rightarrow 0
(t\rightarrow \infty)$, thus it is reasonable to suppose
$p_i(t)\ll 1/N$ when $t$ is sufficiently large. In this case, at
most one agent may change the strategy at each time step (the
probability for two or more agents changing their strategies at
the same time is negligibly small) thus the number of agents on
the majority side is always $(N+1)/2$. Therefore, the agent who
changes the strategy is from the losing side to the losing side
and $p_i(t)$ is reduced by a factor of 2. Since $p_i(t)\ll 1/N$,
the probability that one agent will change his strategy is
$$\eta =1-\prod_{i\in W_l(t)} (1-p_i(t))\approx \sum_{i\in W_l(t)} p_i(t)\approx \frac{G(t)}{2}$$
where $W_l(t)$ is the set of losers at time $t$. Then we have the
iterative equations for $G(t)$ and $H(t)$:
\begin{equation}
G(t+1)=\eta \frac{2N-1}{2N}G(t)+(1-\eta )G(t)
\end{equation}
\begin{equation}
H(t+1)=\eta \frac{H(t)}{2}+(1-\eta )H(t)
\end{equation}
According to {\bf Eq.}(4)\&(5), one can find that $G(t)\sim
t^{-1}$ and $H(t)\sim t^{-N}$\cite{ex3}, which is consentaneous
with the simulation results shown in figure 2.

\begin{figure}
\scalebox{0.45}[0.3]{\includegraphics{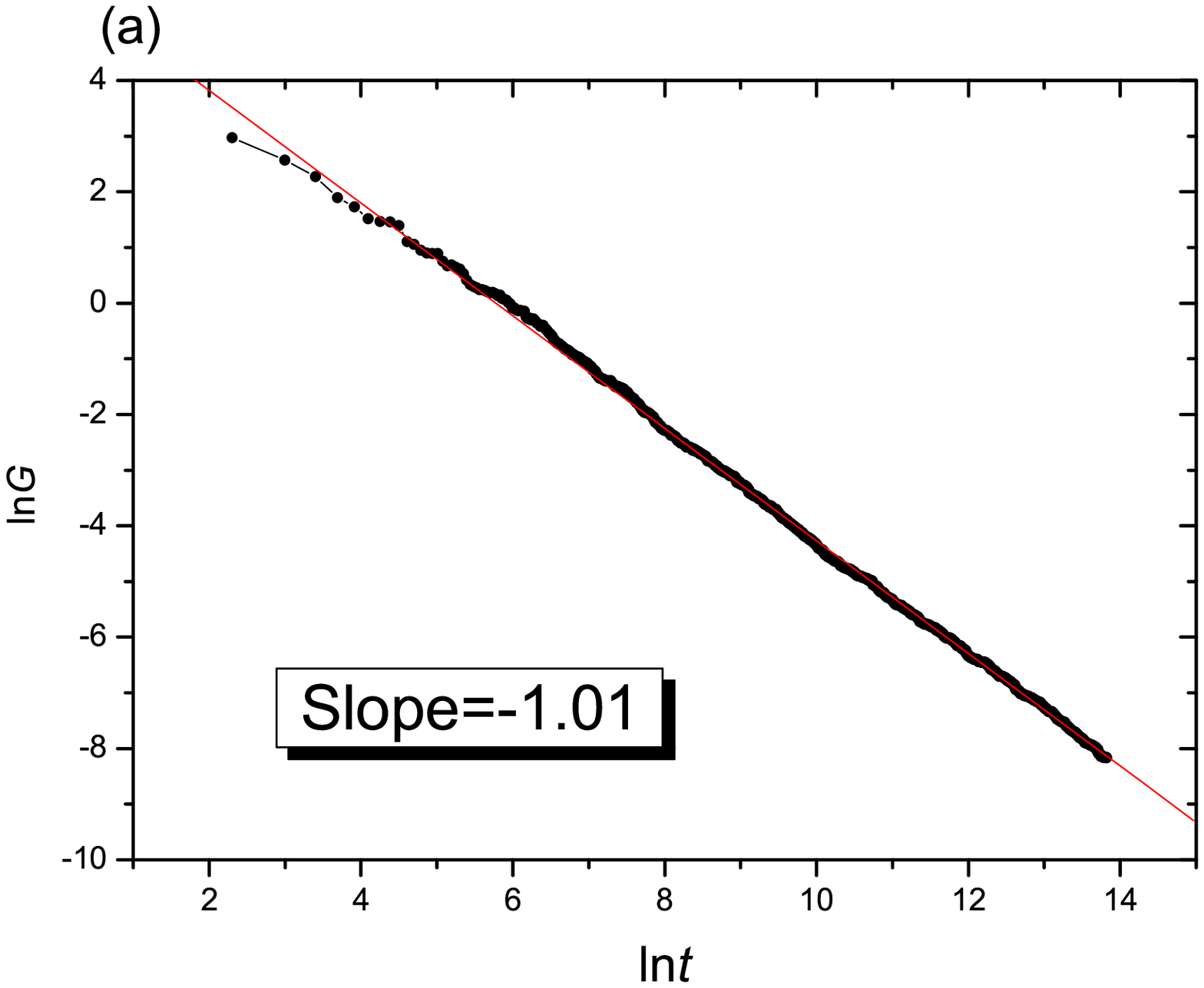}}
\scalebox{0.45}[0.3]{\includegraphics{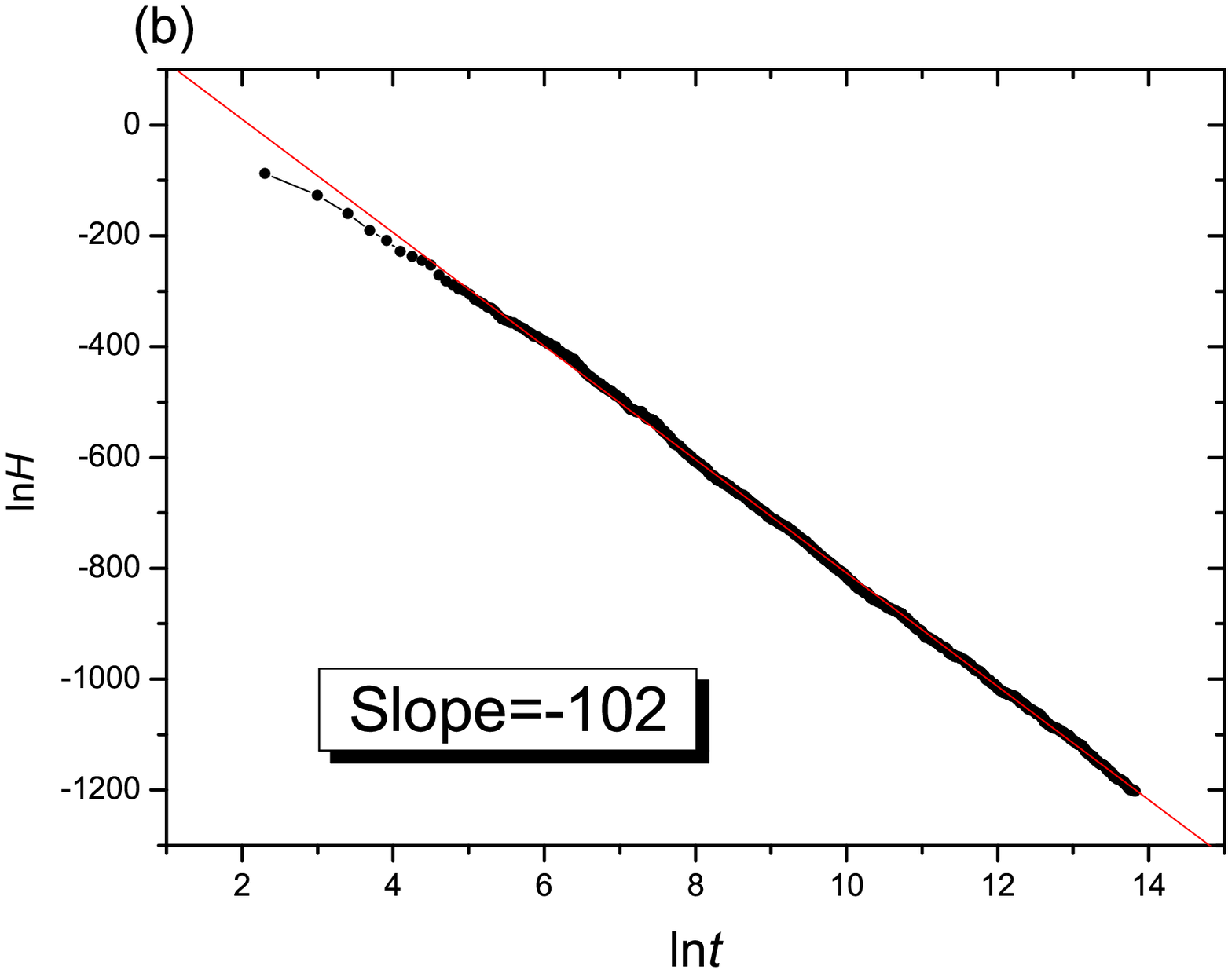}}
\caption{\label{fig:epsart} Time dependence of $G(t)$ (a) and
$H(t)$ (b), where $N=101$ and $m=0$. The slopes of the two curves
in figure {\bf a} \& {\bf b} are $-1.01(\approx -1)$ and
$-102(\approx-N)$, respectively.}
\end{figure}

\section{Smart Agents in Mixed Market}

{\it Challet} et al classified the agents into three different
types\cite{Challet 00}: producers who have only one strategy,
speculators (conventional agents in original minority game) who
have two or more strategies, and the noise traders who make their
choices by random tosses. In this section, we will investigate how
smart agents perform in mixed market\cite{ex4}.

Firstly, let us look into how the smart agents compete with the
producers. Assume that there are $N_p$ producers and $N_s$ $smart$
agents with $N_p+N_s$ an odd integer, each producer has only one
fixed strategy. For simplicity, we shall first discuss the case of
$m=0$. Suppose $N_{p0}$ producers always choose 0, and $N_{p1}$
producers always choose 1. If $\Delta=N_{p0}-N_{p1}>N_s(<-N_s)$,
then $N_s$ smart agents must choose 1(0) in the equilibrium state
and win at each time step. When $N_s>\Delta>0$ (the case
$N_s>-\Delta>0$ is analogic), the situation is slightly
complicated. From the discussion in section 2, it is not difficult
to see that the overall loss of $N_p+N_s$ agents is minimized in
the equilibrium state. Namely, there will be either
$(N_s-\Delta+1)/2$ smart agents choosing 0 and $(N_s+\Delta-1)/2$
smart agents choosing 1 or $(N_s-\Delta-1)/2$ smart agents
choosing 0 and $(N_s+\Delta+1)/2$ smart agents choosing 1. In the
former case, the agents choosing 0 are losers, whiles in the
latter case, the agents choosing 0 are winners. The equilibrium
state is described by the transition between two cases. Before it
switches to another case, the equilibrium state stays in one case
for a period of time, called the life time. The life times of two
cases are different. Assume that the probability $p_i$ of agent
{\bf i} is independent of {\bf i}, then the life time of the
former case is $\tau_1={2/ (N_s-\Delta+1) \langle p\rangle}$ and
the latter case is $\tau_2={2/ (N_s+\Delta+1) \langle p\rangle}$,
where $\langle p\rangle$ denotes the average value of $p_i$. The
overall gain of the smart agents at each time step is equal to
\begin{eqnarray}
\Sigma&=&{1\over \tau_1+\tau_2} [({N_s+\Delta-1\over
2}-{N_s-\Delta+1\over 2})\tau_1
\nonumber\\
&+&({N_s-\Delta-1\over 2}-{N_s+\Delta+1\over 2})\tau_2]
\nonumber\\
&=& {1\over N_s+1}[\Delta^2-1-N_s]
\end{eqnarray}
Therefore, $\Sigma>0$ when $\Delta <N_s<\Delta^2-1$.  The average
profit gained by each smart agent at each time step is
\begin{equation}
{\Sigma\over N_s}={1\over N_s(N_s+1)}[\Delta^2-1-N_s]
\end{equation}

According to {\bf Eq.}(7), when $N_s<\Delta^2-1$, each smart agent
can gain profits from producers. Suppose the number of smart agent
$N_s$ is not fixed, if $N_s<\Delta^2-1$, some new smart agents, if
available, will join the game since they can gain profits from
producers. Thus there will be eventually $N_s\approx \Delta^2-1$
smart agents in the market, whose profits are approximatively
equal to 0 with slight fluctuation. This process can be considered
as an example for the efficient market hypothesis (EMH), which is
hotly controversial in the recent years\cite{EMH}. But in
real-life financial market, the number of producers is alterable,
thus the equilibrium state can rarely be reached.

When $m>0$, the number of possible histories is $2^m>1$. For a
given history $\mu$, suppose $N_{p0}(\mu)$ producers always choose
0 and $N_{p1}(\mu)$ producers always choose 1. Then
$\Delta(\mu)=N_{p0}(\mu)-N_{p1}(\mu)$ is a function of $\mu$.
Since different history $\mu$ is essentially decoupled in our
model and the number of smart agents $N_s$ is fixed, there may be
three cases under history $\mu$: (i) $|\Delta(\mu)|\geq N_s$, each
smart agent can gain one point at each time step; (ii)
$\Delta^2(\mu)-1>N_s>|\Delta(\mu)|$, the smart agents can
averagely gain profit from the producers; (iii)
$\Delta^2(\mu)-1<N_s$, the smart agent cannot gain profit and are
characterized by the overall loss described by {\bf Eq.}(1).

\begin{figure}
\scalebox{0.45}[0.3]{\includegraphics{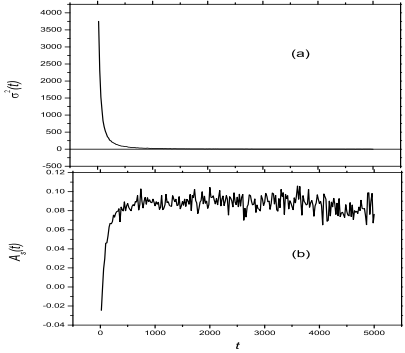}}
\caption{\label{fig:epsart} Time evolution of $\sigma^2 (t)$ (a)
and $A_s(t)$ (b), where $N_p=200$, $N_s=801$, $m=1$ and
$\Delta(0)=\Delta(1)=200$. The value of $\sigma^2 (t)$ and
$A_s(t)$ shown in these two figures is the average of 32
independent experiments and the horizontal line in figure (a)
represents $\sigma^2=0.25$.}
\end{figure}

The above picture is confirmed by the numerical simulation result
shown in Figure 3(a). One can find that $\sigma^2$ decreases as
$t$ increases and decays to 0.25 when $t$ is sufficiently large.
Figure 3(b) plots the time dependence of the mean gain for smart
agents:
$$A_s(t)={N_{s{\rm win}}(t)-N_{s{\rm lose}}(t)\over N_s}$$
where $N_{s{\rm win}}$ and $N_{s{\rm lose}}$ denote the number of
smart agents who win and lose, respectively. Initially, $A_s(t)$
is negative, but as $t$ increases, $A_s(t)$ becomes positive.
Therefore, the smart agents can gain profits from producers in the
regime $\Delta^2(\mu)-1>N_s$.

\begin{figure}
\scalebox{0.45}[0.3]{\includegraphics{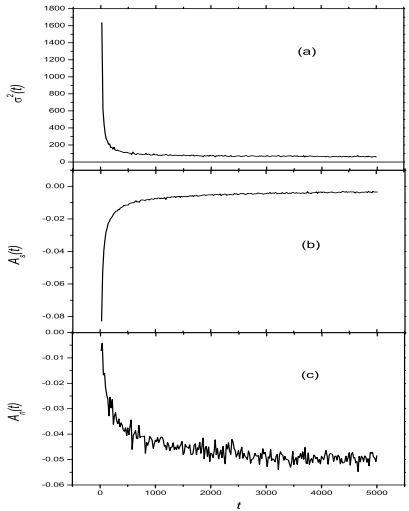}}
\caption{\label{fig:epsart} Time evolution of $\sigma^2 (t)$ (a),
$A_s(t)$ (b), and $A_n(t)$ (c), where $N_s=801$, $N_n=200$ and
$m=1$.The value of $\sigma^2 (t)$, $A_s(t)$ and $A_n(t)$ shown in
these three figures is the average of 32 independent experiments
and the horizontal line in figure (a) represents $\sigma^2=0.25$.}
\end{figure}

Secondly, let's consider the case in which the noise traders and
smart agents are present. Assume that there are $N_n$ noise
traders and $N_s$ smart agents with $N_n+N_s$ an odd integer.
Figure 4(a) plots the time dependence of $\sigma^2$, one can find
that $\sigma^2$ decreases as $t$ increases, but does not reach the
theoretical Optimization 0.25 in the limit of long time. This
result is not difficult to understand for the existence of noise
traders will bring more fluctuations into the system. Figure 4(b)
and 4(c) exhibit the time dependence of $A_s$ and $A_n$
respectively, where $A_n$ is the mean gain of noise traders:
$$A_n(t)={N_{n{\rm win}}(t)-N_{n{\rm lose}}(t)\over N_n}$$
$N_{n{\rm win}}$ and $N_{n{\rm lose}}$ denote the number of the
noise traders who win and lose, respectively. Apparently, the
smart agents perform much better than the noise traders do.

\begin{figure}
\scalebox{0.45}[0.3]{\includegraphics{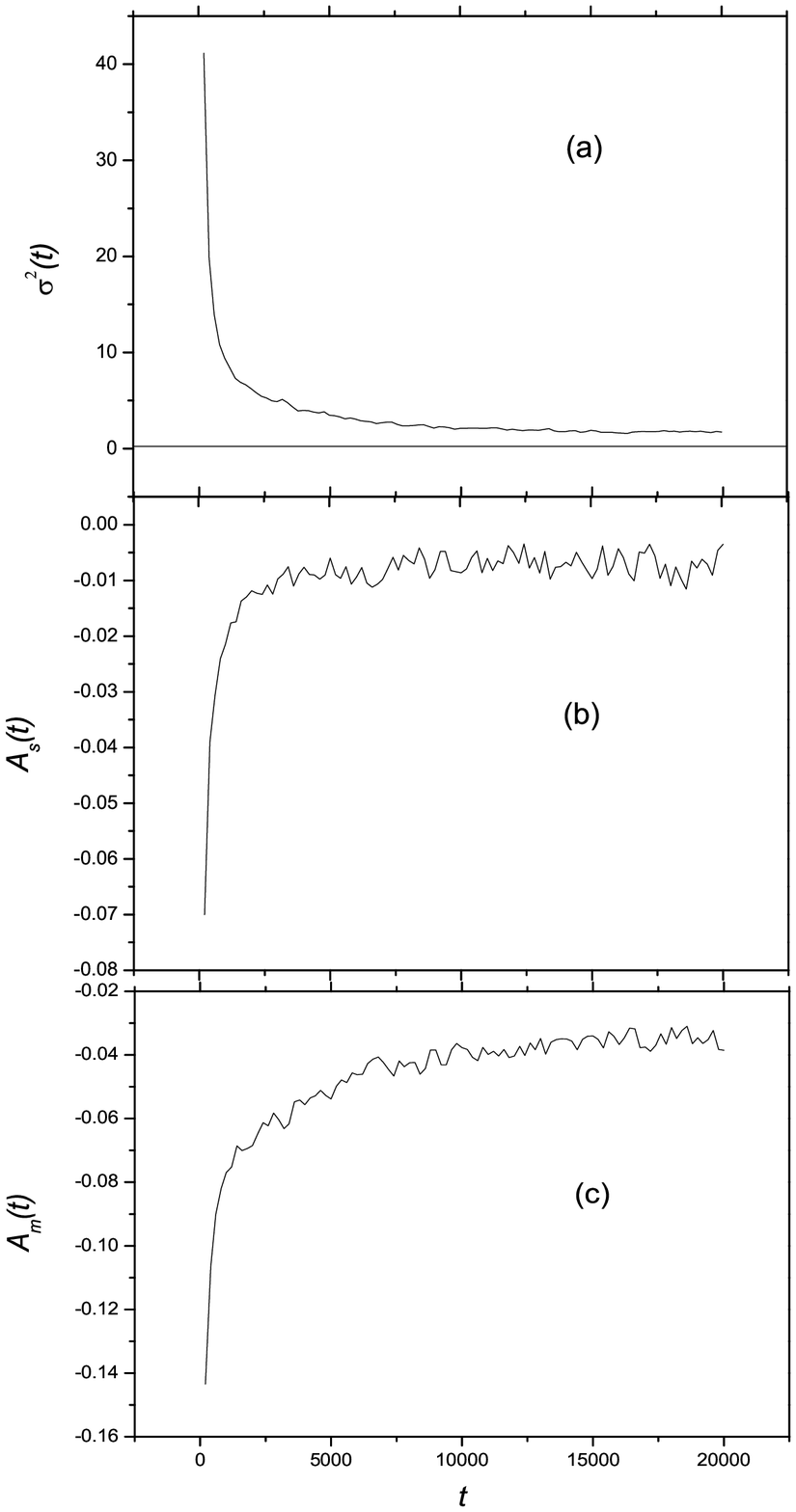}}
\caption{\label{fig:epsart}Time evolution of $\sigma^2 (t)$ (a),
$A_s(t)$ (b), and $A_m(t)$ (c), where $N_s=51$, $N_m=50$, $m=3$
and the number of strategies used by the conventional agents is 2.
The value of $\sigma^2 (t)$, $A_s(t)$ and $A_m(t)$ shown in these
three figures is the average of 32 independent experiments and the
horizontal line in figure (a) represents $\sigma^2=0.25$. }
\end{figure}

At last, We have studied the case in which the conventional
agents, who take the actions based on the original minority game
model\cite{Zhang 97}, and smart agents are present. Assume that
there are $N_s$ smart agents and $N_m$ conventional agents with
$N_s+N_m$ an odd integer. Figure 5(a) shows the time dependence of
$\sigma^2$. One sees that $\sigma^2$ decreases with time but also
does not reach the theoretical Optimization 0.25 in the limit of
long time. This result implies that the conventional agents also
introduce fluctuations, though its magnitude is less than the
noise traders in this case, into the system. In figure 5(b) and
5(c), we report the time dependence of $A_s$ and $A_m$
respectively, where $A_m$ is the mean gain of conventional agents:
$$A_m(t)={N_{m{\rm win}}(t)-N_{n{\rm lose}}(t)\over N_m}$$
where $N_{m{\rm win}}$ and $N_{m{\rm lose}}$ are the number of the
conventional agents who win and lose, respectively. From these two
figures, one immediately finds that the smart agents perform much
better than the conventional agents. This is an evidence that it
may be not reasonable to use the conventional agents to mimic the
actual traders in real-life markets.

\section{Discussion and Conclusion}

We propose a new model of minority game with so-called smart
agents, who use trail and error method to make a choice. When only
the smart agents are present, it is found that the overall loss is
minimized to the theoretical limit as $\sigma^2\to 0.25(t\to\infty
)$. Notice that although those smart agents are independent and
only trying to do their best for their selfish gain based on
inductive thinking, the Global Optimization is achieved in our
model. The result suggests that the economic systems may have the
ability to self-organize into a highly optimized state by agents
who are forced to make decisions based on inductive thinking for
their limited knowledge and capabilities.

In mixed market cases, when the model consists of the smart agents
and the producers with only one fixed strategy, we have found
that, under certain circumstances, the smart agents can gain
profit from the producers.  Also, the overall loss of the
producers and the smart agents is minimized. When the model
consists of the smart agents and the noise traders who choose the
room randomly at each round, it is found that the smart agents
also cooperate very well so that the overall loss of the smart
agents becomes very small when the time is sufficiently large.

It is worthwhile to emphasize that, the smarts agents perform much
better than the conventional agents in mixed market. Imagine an
agent trying to figure out the regularity of the financial market.
Assume at time $t_1$, he has the selection rules for all possible
histories, i.e., he has a strategy. At a later time $t_2$, he
finds that the selection rules for some histories do not give
profits. Therefore, he may change the selection rule for these
history, but not for the other histories which still give him
profits. This is in contrast with the original MG model in which
an agents selects the strategy with the highest virtual point.
When he changes the strategy, he may change many selection rules
although they still make profits. We think that is the reason why
the conventional agents are less competent than smart agents.
\begin{acknowledgements}

This work was supported by the State Key Development Programme of
Basic Research of China (973 Project), the National Natural
Science Foundation of China under Grant No.70271070, the
China-Canada University Industry Partnership Program (CCUIPP-NSFC
No.70142005), the Doctoral Fund from the Ministry of Education of
China, and the National Science Council (Taiwan) under Grant No.
NSC 92-2112-M-001-063.

\end{acknowledgements}


\begin{thebibliography}{Zhang 97}
\bibitem{ex3} In {\bf Eq.}(4), note that $\eta =G(t)/2$, we have
$G^{-2}dG=\frac{1}{4N}dt$, then $\int G^{-2}dG\sim \int
\frac{1}{4N}dt$, and we can obtain the order of $G(t)$: $G(t)\sim
t^{-1}$. Analogously, we have $\int \frac{dH}{H}\sim \int
(\frac{N}{t}+C)dt$, and then $H(t)\sim t^{-N}$, where $C$ is a
constant.
\end{thebibliography}
\end{document}